\documentclass[a4paper,11pt]{article}
\pdfoutput=1 

\usepackage{jinstpub} 
\usepackage{easyReview}

\usepackage{lineno}

\title{\boldmath New technique of ion identification in Accelerator Mass Spectrometry using low-pressure TPC with GEM readout}

\author[a,b]{A. Bondar,}
\author[a,b]{A. Buzulutskov,}
\author[a,b]{V. Parkhomchuk,}
\author[a,b]{A. Petrozhitsky,}
\author[a,b,1]{T. Shakirova, \note{Corresponding author.}}
\author[a,b]{A. Sokolov}

\affiliation[a]{Budker Institute of Nuclear Physics,\\11, Acad. Lavrentieva Pr., Novosibirsk, 630090, Russia}
\affiliation[b]{Novosibirsk State University,\\1, Pirogova str, Novosibirsk, 630090, Russia}

\emailAdd{T.M.Shakirova@inp.nsk.su}

\abstract{We have developed and successfully tested a low-pressure Time Projection Chamber (TPC) with Gas Electron Multiplier (GEM) readout for Accelerator Mass Spectrometry (AMS).

AMS facility in Novosibirsk has a problem of separating isobar ions of different chemical elements that have the same atomic mass. The typical example is radioactive isotopes $^{10}$Be and $^{10}$B that are used to date geological objects at a time scale of ten million years. To solve this problem, a new ion identification technique, namely that based on measuring both ion track ranges and ion energies in low-pressure TPCs with GEM readout, has been developed. This technique is proposed to be applied in AMS for dating geological objects, namely for geochronology of Cenozoic era. 

In this work, we developed a new larger version of the TPC with a dedicated thin silicon nitride window for an efficient passage of ions. The TPC characteristics were studied in isobutane at low pressures using alpha particle sources. In addition, the use of GEM instead of THGEM has been shown to substantially improve the energy resolution at a nominal pressure (50 torr). Using these results and SRIM code simulations, it is shown that isobaric boron and beryllium ions can be effectively separated at AMS, providing efficient dating on a scale of ten million years. This technique will be applied in the AMS facility in Novosibirsk in the near future.}


\proceeding{The 7$^{\text{th}}$ International Conference on Micro Pattern Gaseous Detectors\\
  December 11-16, 2022\\
  Weizmann Institute of Science, Rehovot, Israel}

\begin{document}
\maketitle
\flushbottom

\section{Introduction}
\label{sec:intro}
Cosmogenic radionuclides are rare isotopes formed in nuclear reactions as a result of the irradiation by cosmic rays. The most interesting to study are radioactive isotopes with a long half-life such as $^{10}$Be, $^{14}$C, $^{26}$Al, $^{36}$Cl, $^{41}$Ca and $^{129}$I, which are used in most sciences: archeology, medicine, geology, ecology, etc.~\cite{cosmog_isotopes}. The relative abundance of these isotopes is usually less than 10$^{-12}$ which requires precise separation techniques. Accelerator Mass Spectrometry (AMS) allows to accurately measure fraction of cosmogenic isotopes at the level of 10$^{-15}$ of the total content of the element. The AMS technique is based on the extraction of individual atoms from a sample with subsequent counting of the isotopes of interest. One of the most common examples of the use of AMS is the radiocarbon dating. Using AMS, the radiocarbon age of a small sample of less than 50,000 years can be determined with an accuracy of 0.5\% within a few minutes.

There are approximately 140 AMSs worldwide, one of which was developed at the Budker Institute of Nuclear Physics (BINP). At present, radiocarbon dating is being carried out at the BINP AMS. One of the limitations of the AMS BINP is the isobaric background. Isobars are nuclides of different chemical elements that have the same mass number, such as $^{10}$Be and $^{10}$B. Beryllium-10 is of particular importance in the environmental sciences (archaeology, glaciology, oceanography) and geology for dating purposes in the time interval from 1 thousand to 10 million years. Thus the separation of isobaric ions at the output of AMS is an important problem.

To solve the problem of isobar separation, we have proposed a new technique for ion identification based on measuring the ion ranges and their energies in a low-pressure Time Projection Chamber (TPC) with amplification of the charge signal using a gas electron multiplier (GEM). In previous works~\cite{LPTPC_1,LPTPC_2} we have shown the possibility of separating ions in a low-pressure TPC with THGEM readout in isobutane at a pressure of 120 Torr. In this work, we have developed a large version of the TPC with a special thin silicon nitride window for the passage of ions and placed the chamber inside a vacuum volume. The low-pressure TPC characteristics have been studied in isobutane at a pressure of 50 Torr planned for the operation at the AMS using alpha particles of various energies. In addition, the use of GEM instead of THGEM has been shown to substantially improve the energy resolution. 

\section{Range distribution from SRIM}
\label{sec:SRIM}
\begin{figure}[htbp]
	\centering
	\begin{minipage}{0.45\textwidth}
		\centering
		\includegraphics[width=0.9\textwidth]{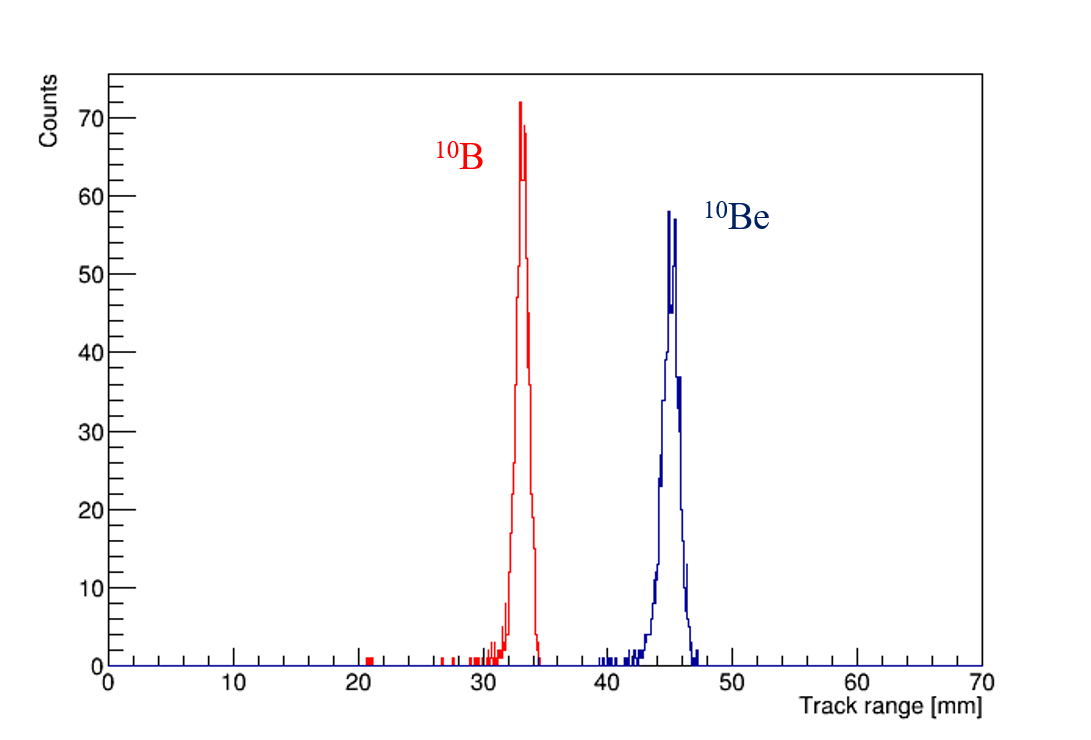}
		\caption{Track range distribution of 4 MeV $^{10}$Be and $^{10}$B passing through 200 nm silicon nitride window into 50 torr isobutane at room temperature. The ranges were obtained using SRIM simulation~\cite{srim}.}		
		\label{fig:SRIM_isobars}
	\end{minipage}\hfill
	\begin{minipage}{0.45\textwidth}
		\centering
		\includegraphics[width=0.9\textwidth]{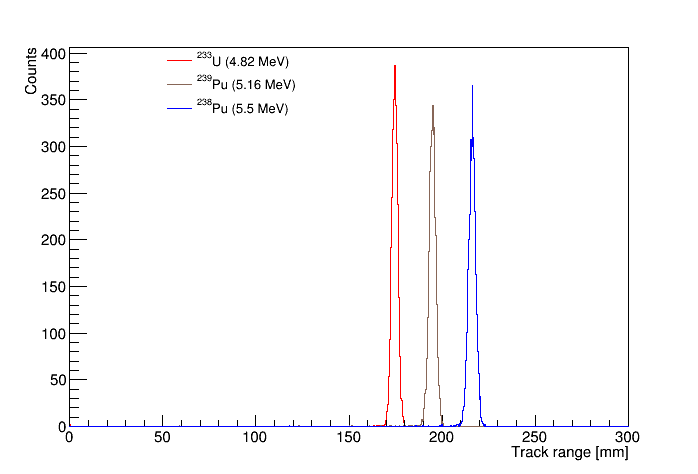}
		\caption{Track range distribution of alpha particles with different energies passing through 200 nm silicon nitride window into 50 torr isobutane at room temperature. The ranges were obtained using SRIM simulation~\cite{srim}.}
		\label{fig:SRIM_alphas}
	\end{minipage}
\end{figure}

The concept of identifying isobaric ions is to separate them by their track ranges. This is possible because ions with different atomic numbers have different energy losses in matter. The SRIM software package~\cite{srim} was used to determine the possibility of separating the $^{10}$Be and $^{10}$B isobars according to the track ranges. The results of track range simulations in isobutane at a pressure of 50 Torr, taking into account the entrance window of silicon nitride with a thickness of 200 nm, are shown in the figure~\ref{fig:SRIM_isobars}. It can be seen that the ion track ranges differ by about 12 mm. Therefore, the low-pressure TPC with a spatial resolution of the order of 2-3 mm will be sufficient to separate such ions.

To determine the spatial resolution of the TPC, the detector was calibrated using a triple source of alpha particles $^{233}$U, $^{239}$Pu and $^{238}$Pu with energies of 4.8 MeV, 5.2 MeV and 5.5 MeV, respectively. The track ranges in isobutane from alpha particles were also simulated under the same conditions as in the experiment. The results are shown in the figure~\ref{fig:SRIM_alphas}.

\section{Experimental setup}
\label{sec:setup}
\begin{figure}[htbp]
	\centering
	\begin{minipage}{0.45\textwidth}
		\centering
		\includegraphics[width=0.9\textwidth]{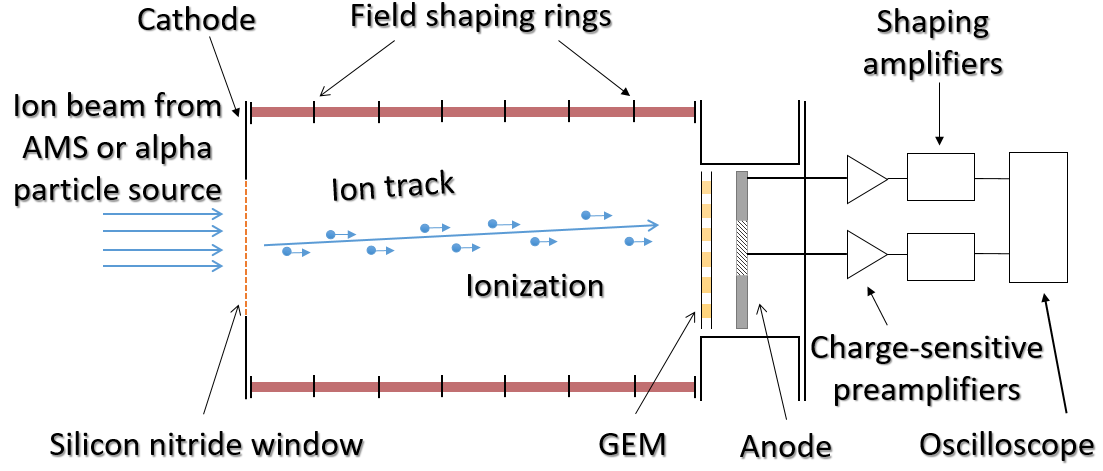}
		\caption{Scheme of a low-pressure time projection chamber (not to scale).}
		\label{fig:exp_setup}
	\end{minipage}\hfill
	\begin{minipage}{0.45\textwidth}
		\centering
		\includegraphics[width=0.9\textwidth]{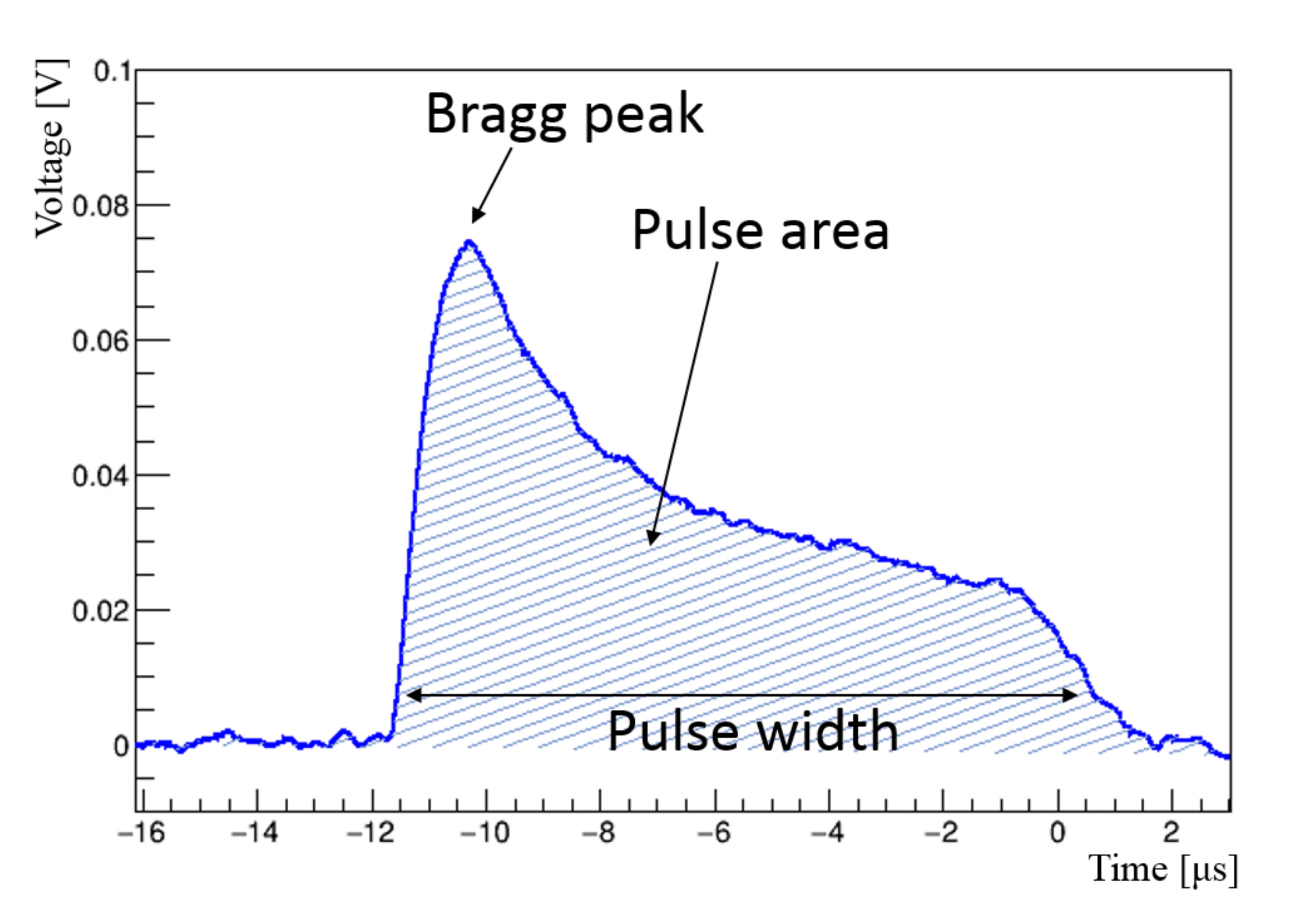}
		\caption{An example waveform from alpha particle in the low-pressure TPC.}
		\label{fig:waveform}
	\end{minipage}
\end{figure}

To calibrate the detector with alpha particles at a nominal operating pressure of 50 Torr, a large version of the low-pressure TPC was developed. A schematic representation of the detector is shown in figure~\ref{fig:exp_setup}. 

The chamber is a 30 cm long ceramic cylinder which is divided by copper field-shaping rings. The internal diameter of the chamber is 10 cm. As the appropriate potentials are applied to these rings, a uniform electric field is created in the detector volume. The volume is filled with isobutane at room temperature and 50 Torr pressure. One flange has a silicon nitride inlet window manufactured by Silson~\cite{silson}. The Si$_3$N$_4$ membrane 10x10 mm$^2$ area and 200 nm in thickness provides the passage for an ion beam from the AMS with energy of 4 MeV or for alpha particles from the radioactive source. These membranes have high strength, low energy loss and less energy straggling compared to plastic entrance window~\cite{membrane}. On the opposite flange, THGEM or GEM are installed on nylon-6 rods, followed by an anode.

Alpha particles or ions from the AMS create primary ionization in the detector volume passing through the gas. The resulting electrons drift towards the GEM or THGEM, where avalanche multiplication takes place. After that induced signal is read out at the anode and then it is amplified and digitized. Figure~\ref{fig:waveform} shows the typical waveform from an alpha particle. When analyzing the data, the pulse width and area were determined, the former being proportional to the ion track range, and the latter being proportional to the deposited energy. The pulse width was calculated at a fixed fraction of the maximum, which was 5\% of the Bragg peak.

\section{Results}
\label{sec:results}
During the operation of the low-pressure TPC with amplification of the charge signal using a THGEM at a pressure of 50 Torr, a charging effect was observed: the signal amplitude decreased with time until it reached a plateau within about two hours. Therefore, it was decided to install GEM instead of THGEM as the charging effect was not found in the former.

\begin{figure}[htbp]
	\centering
	\begin{minipage}{0.45\textwidth}
		\centering
		\includegraphics[width=0.9\textwidth]{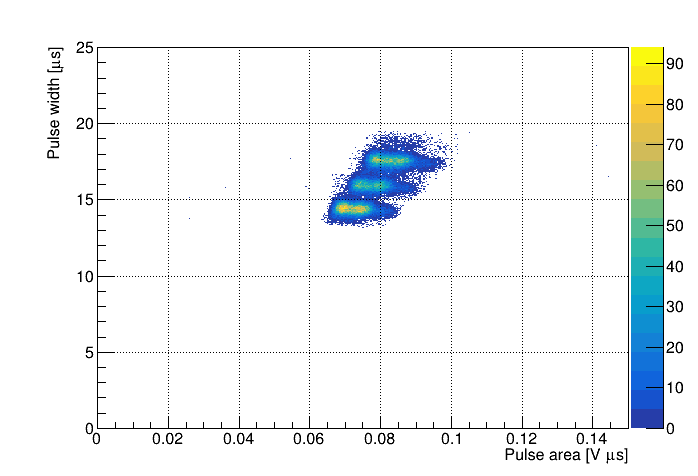}
		\caption{2D plot of pulse width versus pulse area for alpha particles from  $^{233}$U (4.8 MeV),  $^{239}$Pu (5.2 MeV) and  $^{238}$Pu (5.5 MeV) source, measured in low-pressure TPC in isobutane at 50 torr and room temperature using THGEM amplification with gain of 320.}
		\label{fig:2D_plots_THGEM}
	\end{minipage}\hfill
	\begin{minipage}{0.45\textwidth}
		\centering
		\includegraphics[width=0.9\textwidth]{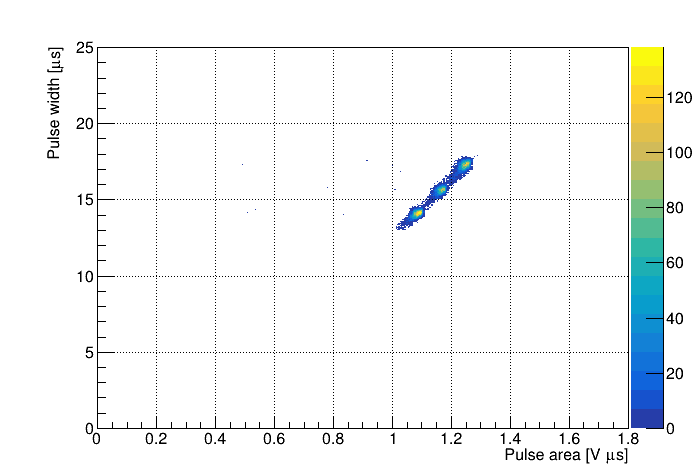}
		\caption{2D plot of pulse width versus pulse area for alpha particles from  $^{233}$U (4.8 MeV),  $^{239}$Pu (5.2 MeV) and $^{238}$Pu (5.5 MeV) source, measured in low-pressure TPC in isobutane at 50 torr and room temperature using GEM amplification with gain of 230.}
		\label{fig:2D_plots_GEM}
	\end{minipage}
\end{figure}

The spectrum of a triple alpha particle source was measured in the low-pressure TPC with both THGEM and GEM. A two-dimensional distribution of the pulse widths and areas is shown in figures~\ref{fig:2D_plots_THGEM} and~\ref{fig:2D_plots_GEM}. Figure~\ref{fig:2D_plots_THGEM} shows the results obtained when working with THGEM, where three areas are visible corresponding to three energies of alpha particles. These areas are well separated by the pulse width and are not separated by the pulse area. However, these regions are well separated both in pulse width and pulse area when using GEM, as shown in figure~\ref{fig:2D_plots_GEM}. It can be seen that the use of GEM instead of THGEM significantly improves the energy resolution at nominal pressure (50 Torr).

\section{Conclusion}
\label{sec:concl}
We have developed and successfully tested a low-pressure TPC with GEM readout for application in AMS. A new ion identification technique that is based on measuring both ion track ranges and ion energies in low pressure TPCs has been developed. Using these results and SRIM code simulations, it is shown that spatial resolution of the TPC amounted to 2.4 mm and thus isobaric boron and beryllium ions can be effectively separated at AMS at a level of 5 sigmas, providing efficient dating on a scale of 10 million years. This technique will be applied in the AMS facility in Novosibirsk in the near future for dating geological objects, in particular for geochronology of Cenozoic Era.

\acknowledgments

This work was supported in part by Russian Science Foundation (project no. 23-22-00359).


\end{document}